\documentclass[conference,onecolumn]{IEEEtran}
\usepackage{bm}
\usepackage{cases}
\usepackage[noadjust]{cite}

\newtheorem{theorem}{Theorem}
\newtheorem{corollary}{Corollary}
\newtheorem{remark}{Remark}

\newcommand{\eqref}[1]{\textnormal{(\ref{#1})}}
\newcommand{\relvar}[2]{\buildrel {#2} \over {#1}}
\newcommand{\eqvar}[1]{\relvar{=}{#1}}

\newcommand{\levar}[1]{\relvar{\le}{#1}}

\begin{document}

\title{Some Extensions of Gallager's Method to General Sources and Channels}

\author{\authorblockN{Shengtian Yang and Peiliang Qiu}
\authorblockA{Department of Information Science \& Electronic Engineering\\
Zhejiang University\\
Hangzhou 310027, China\\
Email: yangshengtian@zju.edu.cn and qiupl@zju.edu.cn}}

\maketitle

\begin{abstract}
The Gallager bound is well known in the area of channel coding. However, most discussions about it mainly focus on its applications to memoryless channels. We show in this paper that the bounds obtained by Gallager's method are very tight even for general sources and channels that are defined in the information-spectrum theory. Our method is mainly based on the estimations of error exponents in those bounds, and by these estimations we proved the direct part of the Slepian-Wolf theorem and channel coding theorem for general sources and channels.
\end{abstract}

\section{Introduction}
In his paper \cite{EMG:Gallager196501} in 1965, Gallager developed a simple inequality technique to derive the coding theorem for memoryless channels without resorting to the law of large numbers. Its central idea may be summarized as the following two basic inequalities.
$$
1\{x' \ge x\} \le \left( \frac{x'}{x} \right)^s, \quad \mbox{$x > 0$, $x' \ge 0$, $s > 0$}
$$
$$
\min\{x, 1\} \le x^\rho, \quad \mbox{$x \ge 0$, $0 < \rho \le 1$}
$$
Although this method is also applicable to channels with memory, Gallager and other researchers mainly concentrated on its applications to memoryless channels. To the best of our knowledge, no one has investigated the extensions of  Gallager's method to general channels so far.

Recently, inspired by Gallager's method and its development \cite{EMG:Gallager196501,EMG:Shulman199909,EMG:Bennatan200403}, we derived a similar upper bound on the average probability of maximum a posterior (MAP) decoding error of Slepian-Wolf codes, and we proved the direct part of Slepian-Wolf theorem for general sources \cite{EMG:Yang200406, EMG:Yang200412}. Compared with the result obtained by information-spectrum methods \cite{EMG:Miyake199509}, our proof is slightly weaker since we assume that the alphabets of correlated sources are finite, but it does suggest that Gallager's method may be applicable to general sources and channels defined in the framework of information-spectrum theory \cite{EMG:Han2003}. Following the idea in \cite{EMG:Yang200406}, we will show in this paper that Gallager's method is applicable to general sources and channels.

\section{Definitions and Notations}

A \emph{general source} in the information-spectrum theory \cite{EMG:Han2003} is defined as an infinite sequence
$$
\bm{X} = \{X^n = (X_1^{(n)}, X_2^{(n)}, \cdots, X_n^{(n)})\}_{n=1}^{\infty}
$$
of $n$-dimensional random variables $X^n$ where each component random variable $X_i^{(n)}$ ($1 \le i \le n$) takes values in the alphabet $\mathcal{X}$ (finite or countably infinite). Analogously, we can define the general correlated sources $\bm{X}\bm{Y}$ as an infinite sequence
$$
\bm{XY} = \{X^nY^n = (X_1^{(n)}Y_1^{(n)}, \cdots, X_n^{(n)}Y_n^{(n)})\}_{n=1}^{\infty}
$$
of $n$-dimensional random variables $X^nY^n$ where each component random variable $X_i^{(n)}Y_i^{(n)} \equiv (X_i^{(n)}, Y_i^{(n)})$ ($1 \le i \le n$) takes values in the product alphabet $\mathcal{X} \times \mathcal{Y}$. We denote the sample space and sample sequence of the $n$-dimensional random variables $X^nY^n$, $X^n$ and $Y^n$ by $\mathcal{X}^n \times \mathcal{Y}^n$, $\mathcal{X}^n$, $\mathcal{Y}^n$ and $x^ny^n$, $x^n$, $y^n$ respectively.

Let $W^n = W^n(\cdot|\cdot)$ be an arbitrary conditional probability distribution satisfying
$$
\sum_{y^n \in \mathcal{Y}^n} W^n(y^n|x^n) = 1, \quad \forall x^n \in \mathcal{X}^n
$$
for each $n = 1, 2, \cdots$. We call the sequence $\bm{W} = \{W^n\}_{n=1}^{\infty}$ a \emph{general channel}.

For convenience, we also use the notations $P_{X}(x)$ and $P_{X|Y}(x|y)$ to substitute for $\Pr\{X=x\}$ and $\Pr\{X=x|Y=y\}$ respectively.

\section{Main Results}

Anyone who is familiar with Gallager's method knows that almost all the results in \cite{EMG:Gallager196501} are obtained by analyzing the properties of the function $E_0^{(n)}(\rho, \bm{X})$ ($0 \le \rho \le 1$) defined by
\begin{equation}\label{eq:ChannelExponent0}
-\frac{1}{n} \ln \sum_{y^n \in \mathcal{Y}^n} \left( \sum_{x^n \in \mathcal{X}^n} \!P_{X^n}(x^n) W^n(y^n|x^n)^{\frac{1}{1+\rho}} \right)^{1+\rho}.
\end{equation}
With the assumption that the input and channel are memoryless and stationary, the function \eqref{eq:ChannelExponent0} can be reduced to
$$
-\ln \sum_{y \in \mathcal{Y}} \left( \sum_{x \in \mathcal{X}} P_{X^1}(x) W^1(y|x)^{\frac{1}{1+\rho}} \right)^{1+\rho}.
$$
Then it is irrelevant to $n$, and Gallager used analytic methods to analyze it. However, when we consider a general input source $\bm{X}$ and a general channel $\bm{W}$, the property of the function \eqref{eq:ChannelExponent0} becomes complex since it may change with $n$ and may converge to zero for any $\rho \in (0, 1]$, and hence Gallager's method are not valid any more. To solve it, we adopted a different method based on estimations of the function \eqref{eq:ChannelExponent0}. We proved the following theorem.

\begin{theorem}\label{th:Channel}
Let $\bm{X}$ be a general source and let $\bm{W}$ be a general channel, then for any $0 < \delta < \underline{I}(\bm{X}; \bm{Y})$, there exists a sequence of $\rho_n$ defined by
\begin{equation}\label{eq:ChannelRho}
\rho_n = \min\{\frac{-\frac{1}{2} \ln \epsilon_n(\delta)}{n(\underline{I}(\bm{X}; \bm{Y}) - \delta)}, 1\},
\end{equation}
such that for any $n \ge 1$,
\begin{equation}\label{eq:ChannelEstimation0}
E_0^{(n)}(\rho_n, \bm{X}) \ge \rho_n(\underline{I}(\bm{X}; \bm{Y}) - \delta) - \frac{3\ln2}{n},
\end{equation}
where $\bm{Y}$ is the output of the channel and $\underline{I}(\bm{X}; \bm{Y})$ is the \emph{spectral inf-mutual information rate} defined by
\begin{equation}\label{eq:DefInfI}
\sup\left\{\beta \left| \lim_{n \to \infty} \Pr\bigl\{\frac{1}{n} \ln \frac{W^n(Y^n | X^n)}{P_{Y^n}(Y^n)} < \beta \bigr\} = 0 \right.\right\},
\end{equation}
and
\begin{equation}\label{eq:DefProbForInfI}
\epsilon_n(\delta) \equiv \Pr\bigl\{\frac{1}{n} \ln \frac{W^n(Y^n | X^n)}{P_{Y^n}(Y^n)} < \underline{I}(\bm{X}; \bm{Y}) - \delta \bigr\}.
\end{equation}
\end{theorem}

\begin{proof}
Without loss of generality, we assume that $P_{Y^n}(y^n) > 0$ for all $y^n \in \mathcal{Y}^n$, and we define
\begin{IEEEeqnarray}{l}
A(y^n, \delta) = \left\{ x^n \in \mathcal{X}^n \left| \frac{1}{n} \ln \frac{W^n(y^n | x^n)}{P_{Y^n}(y^n)} \ge \underline{I}(\bm{X}; \bm{Y}) - \delta \right.\right\}, \label{eq:ChannelA}
\end{IEEEeqnarray}
hence it follows from the definition \eqref{eq:DefInfI} and \eqref{eq:DefProbForInfI} that
$$
\sum_{y^n \in \mathcal{Y}^n} P_{X^nY^n}(A(y^n, \delta)^c, y^n) = \epsilon_n(\delta), \quad \lim_{n \to \infty} \epsilon_n(\delta) = 0.
$$
We further define
\begin{equation}\label{eq:ChannelB}
B(\delta) = \left\{ y^n \in \mathcal{Y}^n \left| P_{X^n|Y^n}(A(y^n, \delta)^c|y^n) \!<\! \epsilon_n(\delta)^{\frac{1}{2}} \right.\right\},
\end{equation}
and
\begin{IEEEeqnarray}{rCl}
P_{Y^n}(B(\delta)^c) &= &\sum_{y^n \in B(\delta)^c} P_{Y^n}(y^n) \IEEEnonumber \\
&\le &\sum_{y^n \in B(\delta)^c} \frac{P_{X^nY^n}(A(y^n, \delta)^c, y^n)}{\epsilon_n(\delta)^{\frac{1}{2}}} \IEEEnonumber \\
&\le &\frac{\sum_{y^n \in \mathcal{Y}^n} P_{X^nY^n}(A(y^n, \delta)^c, y^n)}{\epsilon_n(\delta)^{\frac{1}{2}}} \IEEEnonumber \\
&= &\epsilon_n(\delta)^{\frac{1}{2}}. \label{eq:ChannelBc}
\end{IEEEeqnarray}
Then we have
\begin{IEEEeqnarray*}{Cl}
&\exp\{-nE_0^{(n)}(\rho_n, \bm{X})\} \\
= &\sum_{y^n \in B(\delta)^c} \biggl( \sum_{x^n \in \mathcal{X}^n} P_{X^n}(x^n)^{\frac{\rho_n}{1+\rho_n}} P_{X^nY^n}(x^ny^n)^{\frac{1}{1+\rho_n}} \biggr)^{1+\rho_n} \\
&+ \sum_{y^n \in B(\delta)} P_{Y^n}(y^n) \biggl( \sum_{x^n \in \mathcal{X}^n} P_{X^n}(x^n)^{\frac{\rho_n}{1+\rho_n}} P_{X^n|Y^n}(x^n|y^n)^{\frac{1}{1+\rho_n}} \biggr)^{1+\rho_n} \\
\levar{(a)} &\sum_{y^n \in B(\delta)^c} P_{Y^n}(y^n) + \sum_{y^n \in B(\delta)} P_{Y^n}(y^n) \biggl[ \sum_{x^n \in A(y^n, \delta)^c} P_{X^n}(x^n)^{\frac{\rho_n}{1+\rho_n}} P_{X^n|Y^n}(x^n|y^n)^{\frac{1}{1+\rho_n}} \:+ \\
&\sum_{x^n \in A(y^n, \delta)} P_{X^n|Y^n}(x^n|y^n) \bigl( \frac{P_{X^n}(x^n)}{P_{X^n|Y^n}(x^n|y^n)} \bigr)^{\frac{\rho_n}{1+\rho_n}} \biggr]^{1+\rho_n} \\
\levar{(b)} &\epsilon_n(\delta)^{\frac{1}{2}} + \sum_{y^n \in B(\delta)} P_{Y^n}(y^n) \biggl( P_{X^n|Y^n}(A(y^n, \delta)^c|y^n)^{\frac{1}{1+\rho_n}} + e^{-\frac{n\rho_n}{1+\rho_n} (\underline{I}(\bm{X}; \bm{Y}) - \delta)} \biggr)^{1+\rho_n} \\
\levar{(c)} &\sum_{y^n \in B(\delta)} P_{Y^n}(y^n) \biggl( \epsilon_n(\delta)^{\frac{1}{2(1+\rho_n)}} + e^{-\frac{n\rho_n(\underline{I}(\bm{X}; \bm{Y}) - \delta)}{1+\rho_n}} \biggr)^{1+\rho_n} + \epsilon_n(\delta)^{\frac{1}{2}} \\
\levar{(d)} &e^{-n\rho_n(\underline{I}(\bm{X}; \bm{Y}) - \delta) + 2\ln2} + \epsilon_n(\delta)^{\frac{1}{2}} \\
\levar{(e)} &e^{-n\rho_n(\underline{I}(\bm{X}; \bm{Y}) - \delta) + 3\ln2},
\end{IEEEeqnarray*}
where (a) follows from H\"{o}lder's inequality, and (b) follows from \eqref{eq:ChannelA}, \eqref{eq:ChannelBc} and H\"{o}lder's inequality, and (c) follows from \eqref{eq:ChannelB}, and (d) and (e) from \eqref{eq:ChannelRho}. This concludes \eqref{eq:ChannelEstimation0}.
\end{proof}

By Theorem \ref{th:Channel}, we can easily prove the direct part of the coding theorem for general channels just by showing the following fact.

\begin{corollary}
Let $\bm{W}$ be a general channel, if the coding rate
\begin{equation}
R < C(\bm{W}) \equiv \sup_{\bm{X}} \underline{I}(\bm{X}; \bm{Y}),
\end{equation}
then the function
\begin{equation}\label{eq:ChannelExponent}
E^{(n)}(R) \equiv \max_{X^n} \max_{0 \le \rho \le 1} \{E_0^{(n)}(\rho, \bm{X}) - \rho R\}
\end{equation}
satisfies $nE^{(n)}(R) \to \infty$ as $n \to \infty$.
\end{corollary}

\begin{proof}
For any rate $R < C(\bm{W})$, there exists a general source $\bm{X}$ such that $R < \underline{I}(\bm{X}; \bm{Y})$. Then by Theorem \ref{th:Channel}, we have for any $0 < \delta < \underline{I}(\bm{X}; \bm{Y})$,
\begin{IEEEeqnarray}{rCl}
nE^{(n)}(R) &\ge &n(E_0^{(n)}(\rho_n, \bm{X}) - \rho_n R) \IEEEnonumber \\
&\ge &n\rho_n(\underline{I}(\bm{X}; \bm{Y}) - \delta - R) - 3\ln2 \label{eq:ChannelEstimation}
\end{IEEEeqnarray}
where $\rho_n$ is defined by \eqref{eq:ChannelRho}. Because $n\rho_n \to \infty$ as $n \to \infty$, the lower bound \eqref{eq:ChannelEstimation} goes to infinity as $n \to \infty$ for sufficiently small $\delta$, and this concludes the corollary.
\end{proof}

\begin{remark}
An intermediate result in the proof of Theorem 1 in \cite{EMG:Gallager196501} is that there exists some codes with rate $R$ such that the average probability of maximum likelihood (ML) decoding error satisfies
$$
P_e^{(n)} \le \exp\{-nE^{(n)}(R)\},
$$
and hence the direct part of the coding theorem for general channels is proved.
\end{remark}

In \cite{EMG:Yang200406}, we obtained a similar upper bound on the average probability of maximum a posterior (MAP) decoding error of Slepian-Wolf codes for general sources. There are three terms in the upper bound, and a typical form of the error exponents of these terms may be formulated as follows.
\begin{equation}\label{eq:SourceExponent}
J^{(n)}(R) = \max_{0 \le \rho \le 1} \{\rho R - J_0^{(n)}(\rho)\},
\end{equation}
where
\begin{equation}\label{eq:SourceExponent0}
J_0^{(n)}(\rho) = \frac{1}{n} \ln \!\sum_{y^n \in \mathcal{Y}^n} \!\left( \sum_{x^n \in \mathcal{X}^n} P_{X^nY^n}(x^ny^n)^{\frac{1}{1+\rho}} \right)^{1+\rho}\!.
\end{equation}
By Corollary \ref{co:Source} to be proved in the sequel, we proved the direct part of the Slepian-Wolf theorem. Later, we found that Gallager had already obtained similar bounds for single sources in \cite[Exercise 5.16]{EMG:Gallager1968}. Of course, he only discussed the properties of the bound for stationary memoryless sources. For general sources with finite alphabets, we have the following theorem on the property of \eqref{eq:SourceExponent0}.

\begin{theorem}\label{th:Source}
Let $\bm{XY}$ be a general correlated sources satisfying $|\mathcal{X}| < \infty$, then for any $\delta > 0$, there exists a sequence of $\rho_n$ defined by
\begin{equation}\label{eq:SourceRho}
\rho_n = \min\{\frac{-\frac{1}{2}\ln\epsilon_n(\delta)}{n(\ln|\mathcal{X}| - \overline{H}(\bm{X}|\bm{Y}) - \delta)}, 1\}
\end{equation}
for $\overline{H}(\bm{X}|\bm{Y}) < \ln|\mathcal{X}| - \delta$ (and $\rho_n = 1$ for $\overline{H}(\bm{X}|\bm{Y}) \ge \ln|\mathcal{X}| - \delta$) such that for any $n \ge 1$,
\begin{equation}\label{eq:SourceEstimation0}
J_0^{(n)}(\rho) \le \rho_n(\overline{H}(\bm{X}|\bm{Y}) + \delta) + \frac{3\ln2}{n},
\end{equation}
where $\overline{H}(\bm{X}|\bm{Y})$ is the \emph{spectral conditional sup-entropy rate} defined by
\begin{equation}\label{eq:DefSupH}
\inf\left\{\alpha \left| \lim_{n \to \infty} \Pr\bigl\{\frac{1}{n} \ln \frac{1}{P_{X^n|Y^n}(X^n|Y^n)} > \alpha\bigl\} = 0 \right.\right\},
\end{equation}
and
\begin{equation}\label{eq:DefProbForSupH}
\epsilon_n(\delta) \equiv \Pr\bigl\{\frac{1}{n} \ln \frac{1}{P_{X^n|Y^n}(X^n|Y^n)} > \overline{H}(\bm{X}|\bm{Y}) + \delta \bigl\}.
\end{equation}
\end{theorem}

\begin{proof}
Without loss of generality, we assume that $P_{Y^n}(y^n) > 0$ for all $y^n \in \mathcal{Y}^n$, and we define
\begin{IEEEeqnarray}{l}
A(y^n, \delta) = \left\{ x^n \in \mathcal{X}^n \left| \frac{1}{n} \ln \frac{1}{P_{X^n|Y^n}(x^n|y^n)} \le \overline{H}(\bm{X}|\bm{Y}) + \delta \right.\right\}, \label{eq:SourceA}
\end{IEEEeqnarray}
hence it follows from the definition \eqref{eq:DefSupH} and \eqref{eq:DefProbForSupH} that
$$
\sum_{y^n \in \mathcal{Y}^n} P_{X^nY^n}(A(y^n, \delta)^c, y^n) = \epsilon_n(\delta), \quad \lim_{n \to \infty} \epsilon_n(\delta) = 0.
$$
Analogous to \eqref{eq:ChannelB} and \eqref{eq:ChannelBc}, we further define
\begin{equation}\label{eq:SourceB}
B(\delta) = \left\{ y^n \!\in\! \mathcal{Y}^n \left| P_{X^n|Y^n}(A(y^n, \delta)^c|y^n) \!<\! \epsilon_n(\delta)^{\frac{1}{2}} \right.\right\},
\end{equation}
and we have
\begin{equation}\label{eq:SourceBc}
P_{Y^n}(B(\delta)^c) \le \epsilon_n(\delta)^{\frac{1}{2}}.
\end{equation}
Then we have
\begin{IEEEeqnarray*}{Cl}
& \exp\{nJ_0^{(n)}(\rho_n)\} \\
= &\sum_{y^n \in B(\delta)} P_{Y^n}(y^n) \biggl( \sum_{x^n \in A(y^n, \delta)} P_{X^n|Y^n}(x^n|y^n)^{\frac{1}{1+\rho_n}} + \sum_{x^n \in A(y^n, \delta)^c} P_{X^n|Y^n}(x^n|y^n)^{\frac{1}{1+\rho_n}} \biggr) ^{1+\rho_n} \\
& + \sum_{y^n \in B(\delta)^c} \biggl( \sum_{x^n \in \mathcal{X}^n} P_{X^nY^n}(x^ny^n)^{\frac{1}{1+\rho_n}} \biggr) ^{1+\rho_n} \\
\levar{(a)} &\sum_{y^n \in B(\delta)} P_{Y^n}(y^n) \biggl( e^{\frac{n\rho_n}{1+\rho_n}(\overline{H}(\bm{X}|\bm{Y}) + \delta)} + |\mathcal{X}|^{\frac{n\rho_n}{1+\rho_n}} P_{X^n|Y^n}(A(y^n, \delta)^c|y^n)^{\frac{1}{1+\rho_n}} \biggr) ^{1+\rho_n} \\
& + \sum_{y^n \in B(\delta)^c} |\mathcal{X}|^{n\rho_n} P_{Y^n}(y^n) \\
\levar{(b)} &|\mathcal{X}|^{n\rho_n} \epsilon_n(\delta)^{\frac{1}{2}} + \sum_{y^n \in B(\delta)} P_{Y^n}(y^n) \biggl( e^{\frac{n\rho_n(\overline{H}(\bm{X}|\bm{Y}) + \delta)}{1+\rho_n}} + |\mathcal{X}|^{\frac{n\rho_n}{1+\rho_n}} \epsilon_n(\delta)^{\frac{1}{2(1+\rho_n)}} \biggr) ^{1+\rho_n} \\
\levar{(c)} &e^{n\rho_n(\overline{H}(\bm{X}|\bm{Y}) + \delta) + 2\ln2} + |\mathcal{X}|^{n\rho_n} \epsilon_n(\delta)^{\frac{1}{2}} \\
\levar{(d)} &e^{n\rho_n(\overline{H}(\bm{X}|\bm{Y}) + \delta) + 3\ln2},
\end{IEEEeqnarray*}
where (a) follows from \eqref{eq:SourceA} and Jensen's inequality (or H\"{o}lder's inequality), and (b) follows from \eqref{eq:SourceB} and \eqref{eq:SourceBc}, and (c) and (d) from \eqref{eq:SourceRho}. This concludes \eqref{eq:SourceEstimation0}.
\end{proof}

In the same way, we have the following corollary.

\begin{corollary}\label{co:Source}
Let $\bm{XY}$ be a general correlated sources satisfying $|\mathcal{X}| < \infty$, if the coding rate
\begin{equation}
R > \overline{H}(\bm{X}|\bm{Y}),
\end{equation}
then the function \eqref{eq:SourceExponent} satisfies $nJ^{(n)}(R) \to \infty$ as $n \to \infty$.
\end{corollary}

\begin{proof}
It follows from Theorem \ref{th:Source} that for any $\delta > 0$,
\begin{IEEEeqnarray}{rCl}
nJ^{(n)}(R) &\ge &n(\rho_n R - J_0^{(n)}(\rho_n)) \IEEEnonumber \\
&\ge &n\rho_n(R - \overline{H}(\bm{X}|\bm{Y}) - \delta) - 3\ln2, \label{eq:SourceEstimation}
\end{IEEEeqnarray}
where $\rho_n$ is defined by \eqref{eq:SourceRho}. Because $n\rho_n \to \infty$ as $n \to \infty$, the lower bound \eqref{eq:SourceEstimation} goes to infinity as $n \to \infty$ for sufficiently small $\delta$, and this concludes the corollary.
\end{proof}

\section{Conclusions and Discussions}

An important fact implied above is that the upper bounds obtained by Gallager's method are very tight even for general sources and channels, and we think that stronger version of Theorem \ref{th:Channel} and \ref{th:Source} may be obtained by more sophisticated methods.

The authors want to emphasize the possibility that there may exist some nontrivial and interesting properties about these error exponents. For example, let us see the derivative of the function \eqref{eq:SourceExponent0}. We have
$$
\frac{dJ_0^{(n)}(\rho)}{d\rho} = \frac{1}{n} H(\bar{X}^n(\rho) | \bar{Y}^n(\rho)),
$$
where the distribution of $\bar{X}^n(\rho)\bar{Y}^n(\rho)$ is defined by
\begin{IEEEeqnarray*}{l}
P_{\bar{X}^n(\rho)\bar{Y}^n(\rho)}(x^ny^n) = \frac{P_{X^nY^n}(x^ny^n)^{\frac{1}{1+\rho}} \bigl( \sum_{\hat{x}^n \in \mathcal{X}^n} P_{X^nY^n}(\hat{x}^ny^n)^{\frac{1}{1+\rho}} \bigr)^{\rho}}{\sum_{\hat{y}^n \in \mathcal{Y}^n} \bigl( \sum_{\hat{x}^n \in \mathcal{X}^n} P_{X^nY^n}(\hat{x}^n\hat{y}^n)^{\frac{1}{1+\rho}} \bigr)^{1+\rho}}.
\end{IEEEeqnarray*}
Clearly, $\bar{X}^n(0)\bar{Y}^n(0) \eqvar{d} X^nY^n$, that is, $\bar{X}^n(0)\bar{Y}^n(0)$ and $X^nY^n$ have the same distribution. Hence for $\frac{1}{n}H(X^n|Y^n) < R < \frac{1}{n}H(\bar{X}^n(1) | \bar{Y}^n(1))$, we have
$$
J^{(n)}(R) = \rho_0R - J_0^{(n)}(\rho_0),
$$
where $\rho_0$ satisfies $\frac{1}{n} H(\bar{X}^n(\rho_0) | \bar{Y}^n(\rho_0)) = R$. (Here, we omit the technical details.) Obviously, when $n$ is fixed, the random variable $\bar{X}^n(\rho_0)\bar{Y}^n(\rho_0)$ is a function of $R$, and hence is a simple curve in the space of $n$-dimensional probability distributions. However, when $n$ is considered, the problem becomes complicated due to the complexity of the general correlated source $\bm{XY}$. Therefore, further investigation of these error exponents is needed.

%\newpage

\section*{Acknowledgment}
% optional entry into table of contents (if used)
%\addcontentsline{toc}{section}{Acknowledgment}
This work was supported in part by the Natural Science Foundation of China under Grant NSFC-60472079 and by the Chinese Specialized Research Fund for the Doctoral Program of Higher Education under Grant 2004-0335099.



% http://www.ctan.org/tex-archive/macros/latex/contrib/supported/IEEEtran/bibtex
%\bibliographystyle{IEEEtran.bst}
\bibliographystyle{IEEEtran} % use IEEEtran.bst style
\bibliography{IEEEabrv,ExtMdGlg}

\begin{thebibliography}{1}
\providecommand{\url}[1]{#1}
\csname url@rmstyle\endcsname
\providecommand{\newblock}{\relax}
\providecommand{\bibinfo}[2]{#2}
\providecommand\BIBentrySTDinterwordspacing{\spaceskip=0pt\relax}
\providecommand\BIBentryALTinterwordstretchfactor{4}
\providecommand\BIBentryALTinterwordspacing{\spaceskip=\fontdimen2\font plus
\BIBentryALTinterwordstretchfactor\fontdimen3\font minus
  \fontdimen4\font\relax}
\providecommand\BIBforeignlanguage[2]{{%
\expandafter\ifx\csname l@#1\endcsname\relax
\typeout{** WARNING: IEEEtran.bst: No hyphenation pattern has been}%
\typeout{** loaded for the language `#1'. Using the pattern for}%
\typeout{** the default language instead.}%
\else
\language=\csname l@#1\endcsname
\fi
#2}}

\bibitem{EMG:Gallager196501}
R.~G. Gallager, ``A simple derivation of the coding theorem and some
  applications,'' \emph{{IEEE} Trans. Inform. Theory}, vol.~11, no.~1, pp.
  3--8, Jan. 1965.

\bibitem{EMG:Shulman199909}
N.~Shulman and M.~Feder, ``Random coding techniques for nonrandom codes,''
  \emph{{IEEE} Trans. Inform. Theory}, vol.~45, no.~6, pp. 2101--2104, Sept.
  1999.

\bibitem{EMG:Bennatan200403}
A.~Bennatan and D.~Burshtein, ``On the application of {LDPC} codes to arbitrary
  discrete-memoryless channels,'' \emph{{IEEE} Trans. Inform. Theory}, vol.~50,
  no.~3, pp. 417--437, Mar. 2004.

\bibitem{EMG:Yang200406}
S.~Yang and P.~Qiu, ``A new proof of the {Slepian-Wolf} theorem and performance
  analysis of non-random codes,'' \emph{Acta Electronica Sinica (in Chinese)},
  submitted.

\bibitem{EMG:Yang200412}
------, ``On the performance of linear {Slepian-Wolf} codes for correlated
  stationary memoryless sources,'' in \emph{Proc. DCC 2005}, to be published.

\bibitem{EMG:Miyake199509}
S.~Miyake and F.~Kanaya, ``Coding theorems on correlated general sources,''
  \emph{IEICE Trans. Fundamentals}, vol. E78-A, no.~9, pp. 1063--1070, Sept.
  1995.

\bibitem{EMG:Han2003}
T.~S. Han, \emph{Information-Spectrum Methods in Information Theory}.\hskip 1em
  plus 0.5em minus 0.4em\relax Berlin: Springer, 2003.

\bibitem{EMG:Gallager1968}
R.~G. Gallager, \emph{Information Theory and Reliable Communications}.\hskip
  1em plus 0.5em minus 0.4em\relax New York: Wiley, 1968.

\end{thebibliography}

\end{document}